\documentclass[twocolumn]{aastex701}
\usepackage{hyperref}
\usepackage{amsmath}
\usepackage[inline]{enumitem}
\usepackage{xcolor}
\usepackage{graphicx}
\usepackage{floatrow}
\usepackage{makecell} % for easy line breaks inside table cells

\begin{document}

\title{Locating Centers of Clusters of Galaxies with Quadruple Images: Witt’s Hyperbola and a New Figure of Merit}

\author{Nixon Hanna}
%\author[0009-0008-1477-4716]{Nixon Hanna}
\affiliation{MIT Department of Physics}
\email{noxin@mit.edu}
% \email[show]{noxin@mit.edu}

\author{Paul L. Schechter}
%\author[0000-0002-5665-4172]{Paul L. Schechter}
\affiliation{MIT Department of Physics}
\affiliation{MIT Kavli Institute for Astrophysics and Space Research}
\email{schech@achernar.mit.edu}

\author{Michael A. McDonald}
%\author[0000-0001-5226-8349] {Michael A. McDonald}
\affiliation{MIT Department of Physics}
\affiliation{MIT Kavli Institute for Astrophysics and Space Research}
\email{mcdonald@space.mit.edu}

\author{Marceau Limousin}
%\author[0001-0001-6636-4999]{Marceau Limousin}
\affiliation{Aix Marseille Univ, CNRS, CNES, LAM, Marseille, France}
\email{marceau.limousin@lam.fr}

\begin{abstract}

For any elliptical potential with an external parallel shear, Witt has proven that the gravitational center lies on a rectangular hyperbola derived from the image positions of a single quadruply lensed object. Moreover, it is predicted that for an isothermal elliptical potential the source position both lies on Witt’s Hyperbola and coincides with the center of Wynne’s Ellipse (fitted through the four images). Thus, by fitting Witt’s Hyperbolae to several quartets of images—ten are known in Abell 1689—the points of intersection provide an estimate for the center for the assumed isothermal elliptical potential. We introduce a new figure of merit defined by the offset of the center of Wynne’s Ellipse from Witt’s Hyperbola. This offset quantifies deviations from an ideal elliptical isothermal potential and serves as a discriminant to exclude poorly fitted quadruples and assign greater weight to intersections of hyperbolae of better fitting systems. Applying the method to 10 quads (after excluding 7 poorly fitted quads) in Abell 1689, we find the potential is centered within 11" of the BCG, X-ray center, flexion-based center and the center found from a total strong lensing analysis. The Wynne-Witt framework thus delivers a fast, analytic, and self-consistency-checked estimator for centers in clusters with multiple quads.

\end{abstract}


\section{Introduction} \label{sec:intro}

\begin{figure}[ht!]
\centering
\includegraphics[scale=0.335]{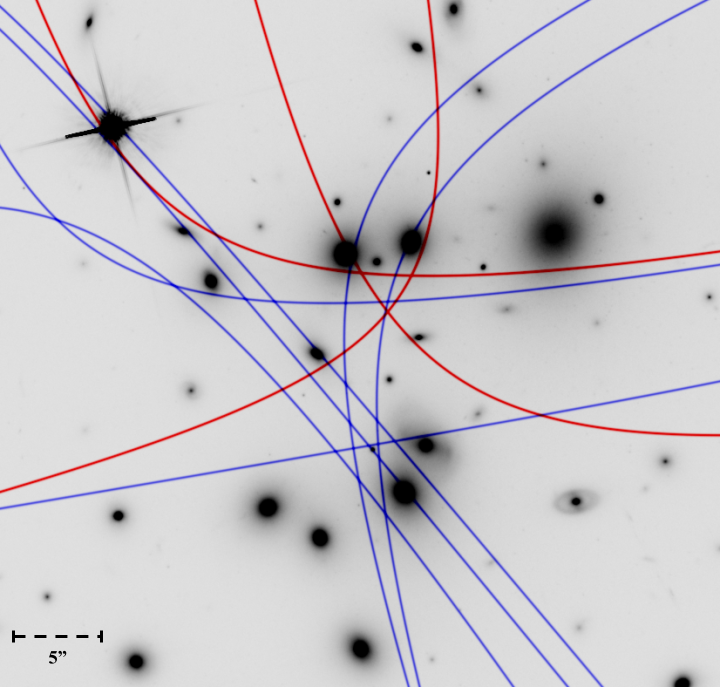}
\caption{Grayscale-inverted HST/ACS F814W view of the core of Abell 1689 with "primary" branches of Witt's Hyperbolae over-plotted. Red lines indicate well-fitting quadruples, while blue lines indicate quadruples that are discarded due to high offsets. The knot where the red hyperbolae intersect is the Wynne–Witt estimate of the cluster’s center of gravitational potential; its small dispersion visually demonstrates the precision of the analytic construction discussed in Sections \ref{sec:method} and \ref{sec:results}.
\label{fig:teaser}}
\end{figure}

In gravitational lensing theory, quadruple gravitational lens configurations admit geometric (as opposed to algebraic) models that tightly constrain the underlying lensing potential. \citet{Witt} demonstrated a fundamental property: for any elliptical gravitational potential with an external parallel shear, the true gravitational center must reside on a rectangular hyperbola—which we call Witt’s Hyperbola—constructed solely from the observed positions of four lensed images of a single background source. Thus, the points of intersection of multiple Witt Hyperbolae provides an estimate for the gravitational center (see Figure~\ref{fig:teaser}). Moreover, \citet{Wynne} showed that for an elliptical isothermal potential, the source position coincides precisely with the center of an ellipse passing through the same four image positions, which we call Wynne’s Ellipse. An elegant combination of Witt’s Hyperbola and Wynne’s Ellipse offers an analytical framework that tightly constrains both the lens center and the source location, contingent on an elliptical isothermal lens potential.

While Witt’s hyperbola, usually supplemented by Wynne’s ellipse, has been applied mostly to galaxy-scale lenses, their potential value at cluster scales is still unexplored. Reliable cluster centers are required not only for precise lens models, but also for multi-wavelength cross-checks. Alternative schemes for determining the centers of clusters include the X-ray surface-brightness peak, the Sunyaev–Zeldovich (SZ) pressure maximum, and the peak of total mass distribution from full strong lensing or flexion analysis. Offsets among these tracers encode important clues to the dynamical state, orientation, and non-thermal pressure support of the intracluster medium \citep{Ettori2020, Shan2010}. In this work we intentionally do not compare our center to weak-lensing centers, for two practical reasons. First, weak-lensing measurements are, almost by definition, obtained well outside the Einstein ring. Second, achieving small uncertainties requires averaging over many background sources, so the resulting centroid is not directly comparable to our strongly constrained center derived from multiply imaged systems.

For the purpose of evaluating self-consistency, we introduce a new figure of merit—the offset between Witt’s Hyperbola and the center of Wynne’s Ellipse—that naturally gauges departures from ideal ellipticity and isothermality. Our proposed offset-based figure of merit provides a fast and quantitatively transparent measure to assess image-set quality and lens potential ellipticity  and isothermality. This measure provides an objective criterion to discriminate against poorly fitted image configurations (as seen in Figure~\ref{fig:teaser}) and prioritizes high-quality quadruples with minimal deviations from ideal potentials.

Abell 1689 provides an especially promising testbed given the prevalence of strongly lensed objects. The Einstein radius of Abell 1689 is exceptionally large, $\theta_{E}\simeq45''$, making it one of the most powerful known gravitational lenses and dramatically increasing the likelihood of locating quadruply imaged systems for our analysis \citep{TuEtal2008,Broadhurst2005}. It is classified as richness class~4 in the original Abell catalogue\footnote{Richness classes 0–5 correspond to increasing galaxy counts; class~4 contains 200–299 galaxies, with Abell 665 being the only class~5 ($>$299 galaxies) cluster}—the second-richest tier—so the cluster contains a dense population of member galaxies in a large region of critical density \citep{Alamo2013}\footnote{More information on critical density, including a definition in Eq. 17, can be found in \citet{Narayan}}. Moreover, Abell 1689 has been the focus of sustained observational effort for more than two decades: a NASA/ADS search on 10~July~2025 returned ${\sim}210$ refereed papers that mention the cluster in their abstracts, underscoring the depth of ancillary data available for cross-checks and follow-up studies. These attributes collectively make Abell 1689 a natural laboratory for assessing the practical utility of Witt’s Hyperbola and Wynne’s Ellipse in complex, cluster-scale lenses. 

In what follows, we detail the mathematical underpinnings and practical implementation of this Wynne-Witt framework (Section \ref{sec:method}). We then illustrate the application of our method using multiple quadruply-lensed image sets from Abell 1689 (Section \ref{sec:results}). Next, we compare our estimated center to other centers and discuss the significance of the offsets (Sections \ref{sec:compare} and \ref{sec:discuss}). Finally, we provide opportunities for future work (Section \ref{sec:future_work}). 

Our results indicate that the Wynne-Witt approach, supplemented by our new figure of merit, provides a robust and computationally efficient estimator for identifying the centers of cluster potentials, which complements alternative methods of measuring those centers.

\section{Conic Sections and Lenses}\label{sec:method}

The analytical foundations underpinning Witt’s Hyperbola and Wynne’s Ellipse are presented in Appendices~\ref{app:hyperbola} and~\ref{app:ellipse}. The salient results of these calculations are summarized in the following sections.

\begin{figure*}[t]
\floatbox[{\capbeside\thisfloatsetup{capbesideposition={left,top},capbesidewidth=5.8cm}}]{figure}[\FBwidth]
{\caption{Grayscale-inverted HST/ACS F814W view of the core of Abell 1689 with zoomed insets marking representative image positions used in our analysis (see Section~\ref{sub:impl} and Section~\ref{sec:results}). Primary branches of Witt's Hyperbolae are plotted in the background image, with insets showing the intersection of Witt's Hyperbolae and Wynne's Ellipses. The conics intersect precisely at the image coordinate so any visual displacement is the result of the dataset. Note: Not all insets have the same magnification.}\label{fig:insets}}
{\includegraphics[width=.53\textwidth]{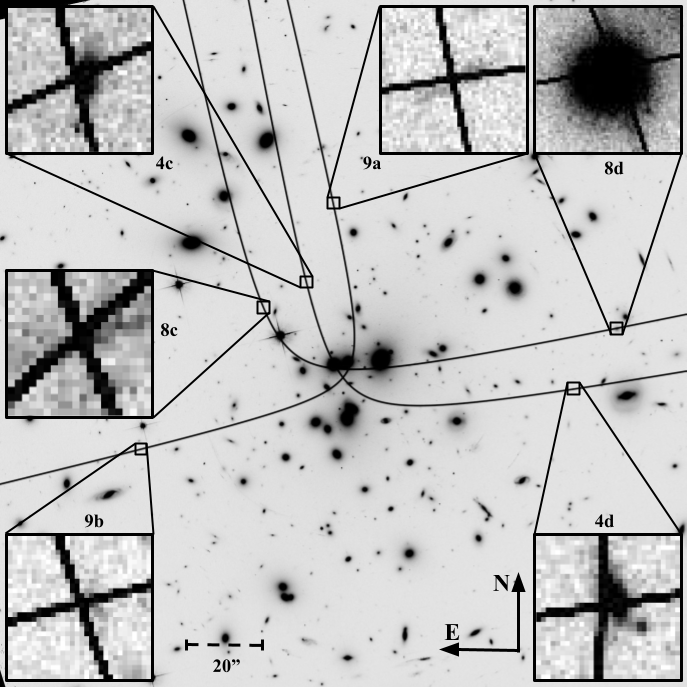}}
\vspace{2em}
\end{figure*}

\subsection{Witt's Hyperbola}

For a gravitational lens whose potential is elliptical with parallel external shear, \citet{Witt} showed that the four observed image positions lie on a \emph{rectangular} hyperbola. The explicit quadratic form assuming the rectangularity condition is given by Eq.~\ref{eq:hyperbola_conic}. While conic sections generically have 5 independent coefficients, the rectangularity of Witt’s hyperbola reduces this to four. We henceforth take the external shear to be identically zero, on the premise that clusters are less likely to be strongly sheared than individual galaxies.

\subsection{Wynne's Ellipse}

\citet{Wynne} further proved that if the potential is isothermal the (unknown) source position coincides with the center of an ellipse that (i) passes through the same four images, (ii) has its principal axes parallel to one of the two asymptotes of Witt’s Hyperbola, and (iii) is centered on the "primary" branch of the hyperbola as discussed in Appendix~\ref{app:ellipse}. The conic section for Wynne’s ellipse has five independent coefficients. All steps in the outlined method involve fixed-size linear or polynomial algebra. The solution for Wynne’s ellipse therefore does not require any iterations and has a constant time complexity, denoted $\mathcal{O}(1)$; the most expensive operation is the root-finding for a degree-6 polynomial, which is still negligible in practice. 

\subsection{Proposed Figure of Merit}

Having relegated the full algebraic derivations to the appendices, we summarize here the operative idea. A quadruply imaged source produces 4 pairs of coordinates, giving eight constraints on the nine conic coefficients for Witt’s hyperbola and Wynne’s ellipse. The alignment of Wynne’s ellipse with Witt’s hyperbola gives a ninth constraint, and its centering on Witt’s hyperbola gives a tenth. The coefficients are therefore overdetermined. Previous treatments of the problem have minimized the distances of the four images from Witt’s hyperbola and Wynne’s ellipse. The alternative adopted in this paper is to absorb all of the mismatch into the offset of the center of the ellipse from the hyperbola. An in-depth comparison of these approaches is given in Appendix~\ref{app:comparison}.

The inverse of this offset is then a figure of merit for the consistency of the Witt-Wynne model with the observed image positions. Let $\mathbf c_{W}$ be the center of Wynne's ellipse and $\mathbf h_{W}$ be the point on Witt's hyperbola that lies \emph{closest} to $\mathbf c_{W}$. We define the offset vector
\begin{equation}
\Delta\mathbf x_{WW}\;\equiv\;\mathbf c_{W}-\mathbf h_{W}\quad.
\end{equation}

Its modulus $|\Delta\mathbf x_{WW}| = \|\Delta\mathbf x_{WW}\|$ measures the offset of Wynne’s hyperbola from the center of Witt’s ellipse. Its inverse serves as our \emph{figure of merit}. Its computation follows the prescription in Appendix~\ref{app:ellipse}. In an exactly elliptical isothermal potential the two conics are perfectly self–consistent, giving $|\Delta\mathbf x_{WW}|=0$. Non–zero values therefore quantify departures arising from substructure, external perturbations, and deviations from isothermality. In practice we adopt a threshold $\epsilon$ of $6\farcs25$ and \emph{discard} any image quartet with $|\Delta\mathbf x_{WW}|>\epsilon$. It is important to note that this threshold is specific to the accidents of this particular cluster and should be expected to be different for other clusters. Calculating this threshold given certain properties of a cluster might be possible but is beyond the scope of the present paper. For the retained sets we optionally weight each hyperbola intersection by $1/|\Delta\mathbf x_{WW}|$, thereby allowing more self–consistent configurations to exert greater influence on the final cluster–center estimate.

\subsection{Practical Implementation} \label{sub:impl}

\begin{enumerate}
\item Fit Witt’s Hyperbola to each image quartet via Eq.~\ref{eq:hyperbola_linear_system}.
\item Compute the finite asymptote slopes $m_{1},m_{2}$ and construct the corresponding candidate ellipses satisfying the five constraints stated above and further discussed in Appendices~\ref{app:hyperbola} and~\ref{app:ellipse}.
\item For each candidate, compute the center $(x_e,y_e)$ and evaluate the polynomial $P(\lambda)$ governing the point-to-hyperbola distance; retain the ellipse with the smaller distance $d$.
\item Record $|\Delta\mathbf x_{WW}|$ from the selected ellipse.
\item Discard configurations with $|\Delta\mathbf x_{WW}|>\epsilon$ (we adopt $\epsilon=6\farcs25$, for Abell 1689) and find the points of intersection of the surviving hyperbolae, optionally weighting intersections by the amplitude of their offset.
\end{enumerate}

\section{Abell 1689 Analysis} \label{sec:results}

We have applied the Witt-Wynne construction to the ten quadruply imaged systems in Abell 1689 identified by \citet{Coe}\footnote{Some objects had more than 4 images due to the splitting of images by substructure in the cluster potential. In these cases, we averaged the positions of the two images and treated this average as the image location.}, following the recipe of Section~\ref{sub:impl}, with result shown in Figure~\ref{fig:insets}. Quantitative details are given in the following subsections. 

\begin{deluxetable}{cccccc}[h]
\tablecaption{Quartet properties and selection.\label{tab:quads}}
\tablehead{
\colhead{ID} &
\colhead{$|\Delta\mathbf x_{WW}|$} &
\colhead{$a/b$} &
\colhead{P.A.} &
\colhead{Distortion} &
\colhead{Status} \\
\colhead{} &
\colhead{(\arcsec)} &
\colhead{} &
\colhead{($^\circ$)} &
\colhead{} &
\colhead{}
}
\startdata
 4  & 0.81  & 1.36 & 100.13 & None & kept     \\
12  & 2.45  & 1.09 &  99.35 & Stretch, 2 Splits & rejected \\
 9  & 3.01  & 1.23 & 103.36 & None & kept     \\
 8  & 6.06  & 1.45 & 102.01 & Stretch & kept     \\
 1  & 6.49  & 1.54 & 109.58 & Split & rejected \\
 2  & 9.34  & 1.58 & 110.60 & Stretch & rejected \\
19  & 9.43  & 1.50 & 101.44 & Stretch & rejected \\
24  & 9.48  & 1.73 & 129.51 & None & rejected \\
29  & 10.99 & 1.60 & 128.84 & None & rejected \\
42  & 11.44 & 1.44 & 122.27 & Stretch & rejected \\
\enddata
\end{deluxetable}

\begin{figure}[t]
\centering
\includegraphics[width=\textwidth]{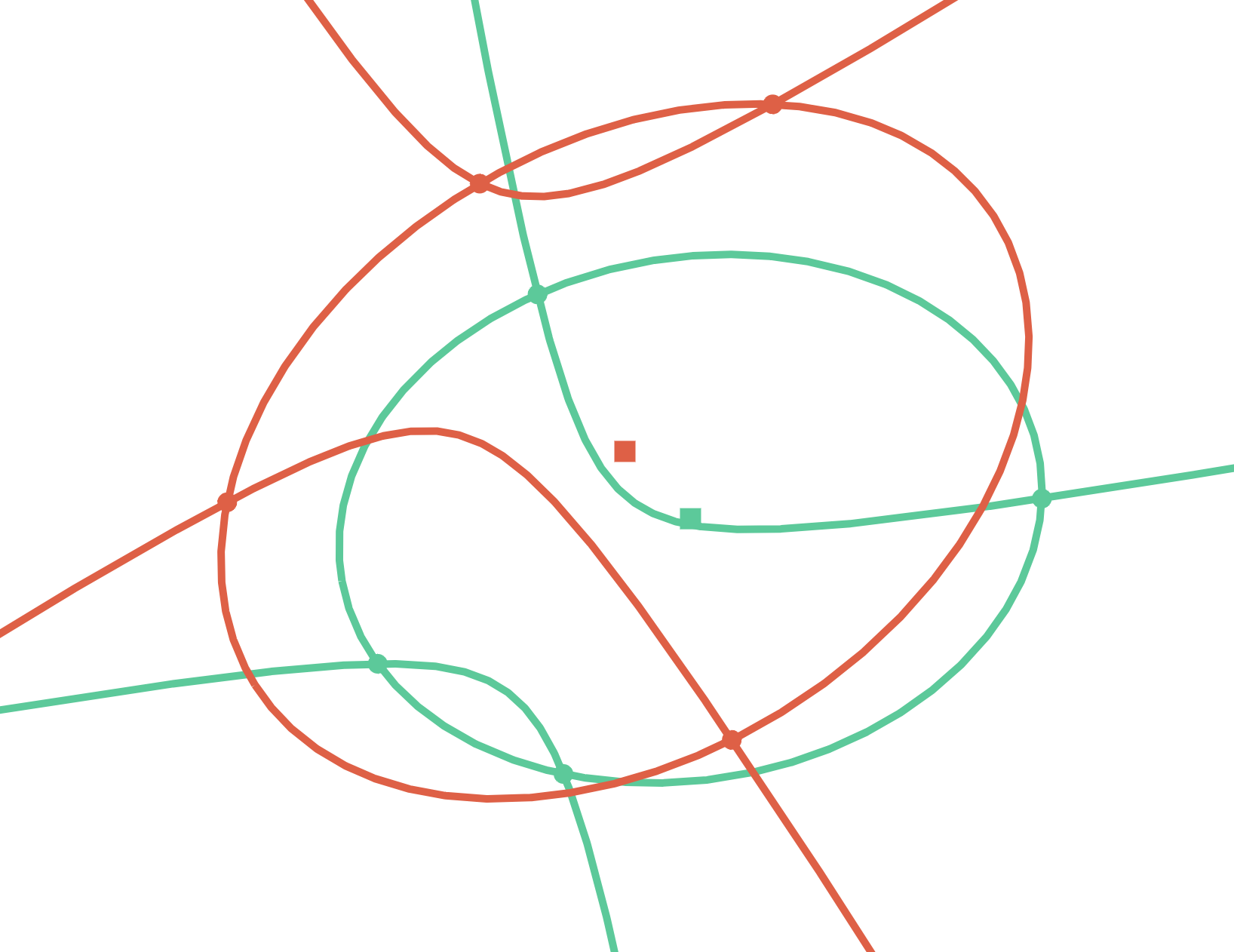}
\caption{Comparison of fit quality between Quad 4 (green) and Quad 42 (red). The green and red squares represent the centers of Wynne's Ellipses for Quad 4 and 42 respectively. Quad 4 can seen to have a very low $|\Delta\mathbf x_{WW}|$, given the extremely small offset between Witt's Hyperbola and the center of Wynne's Ellipse. While Quad 42 is shown to have a larger $|\Delta\mathbf x_{WW}|$, representing a worse model fit.
\label{fig:ellipse_compare}}
\end{figure}

\subsection{Pruning with the offset figure of merit}

\begin{figure*}[t]
\centering
\includegraphics[width=0.49\textwidth]{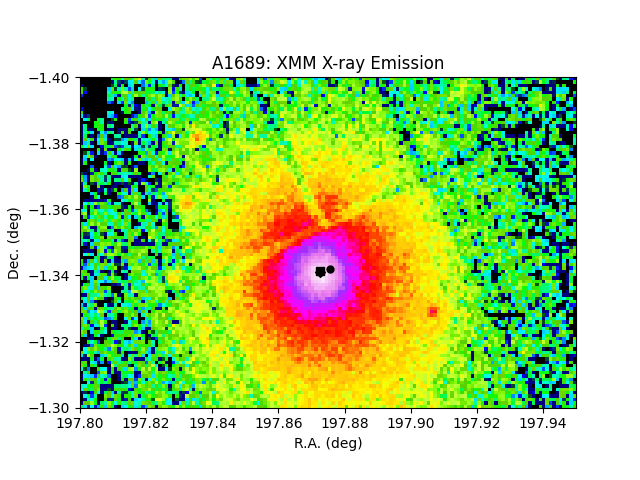}
\includegraphics[width=0.49\textwidth]{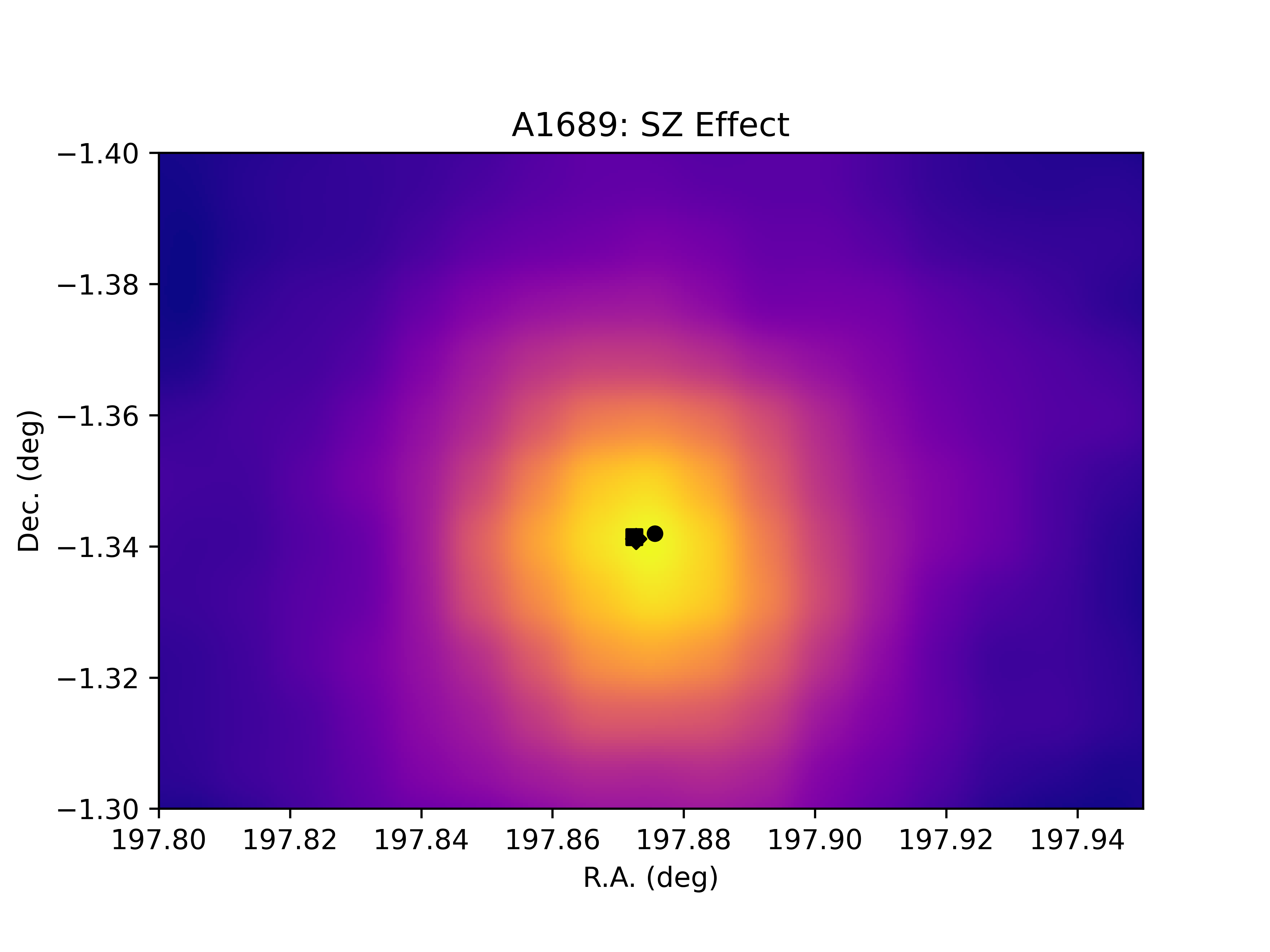}
\caption{$(\alpha,\delta)_{\rm WW}$ (black circle) plotted over x-ray emission and SZ effect data. X-ray center and SZ center are marked with a square and diamond respectively but overlap considerably. X-ray emission data comes from Fig. 2 in \citet{Chappuis}. SZ effect data is a gaussian interpolation of NASA's ACT DR6 + Planck Compton-y Map.}\label{fig:gas_compare}
\end{figure*}

Table~\ref{tab:quads} lists the offsets, axis ratios, position angles, and any extreme distortions that may affect data quality (such as drastically stretched images and images which have been split by cluster members) for all ten quartets. We adopt a conservative offset cutoff of $6\farcs25$ and additionally remove Quad 12, whose fitted ellipse has an axis ratio $a/b=1.09$ (nearly round). In the Wynne construction (see Appendix~\ref{app:ellipse}), the ellipse's principal axes are parallel to the asymptotes of Witt's rectangular hyperbola and it’s axis ratio is equal to the ellipticity of the lens potential. Consequently, if all quartets were probing a single globally elliptical isothermal potential, their amplitude ellipses would share a position angle (P.A.) and a common axis ratio ($a/b$). 

The retained systems (Quads 4, 9, and 8) satisfy both criteria: small offsets and mutually consistent ellipse parameters, with P.A. clustered near $100$--$104^\circ$ and moderate flattenings ($a/b\approx1.2$--$1.5$). By contrast, several rejected cases exhibit large $|\Delta\mathbf x_{WW}|$ together with P.A. and axis ratios that are inconsistent with a single underlying elliptical isothermal potential (e.g., Quads 24 and 29 with $P.A.\simeq130^\circ$ and $a/b\gtrsim1.6$; Table~\ref{tab:quads}). While pruning via $|\Delta\mathbf x_{WW}|$, $a/b$, and P.A. allows us to search for quartets that probe a single isothermal elliptical potential, it is still susceptible to smaller perturbations of image locations by individual cluster members.

\subsection{Intersection of Witt’s hyperbolae}

For the three retained quartets, we now consider the pairwise intersections of the primary branches of their Witt hyperbolae shown in Table~\ref{tab:intersections}. 

\begin{deluxetable}{ccc}[b]
\tablecaption{Points of intersection of primary branches of Witt's Hyperbolae for Quads 4, 8, and 9.\label{tab:intersections}}
\tablehead{
\colhead{IDs} & \colhead{R.A.} & \colhead{Dec.}
}
\startdata
4 \& 8 & 197.87601 & -1.34173 \\
4 \& 9 & 197.87558 & -1.34235 \\
8 \& 9 & 197.87521 & -1.34177 \\
\enddata
\end{deluxetable}

% shown in Table~\ref{tab:intersections}. 

The spread among the three intersections is $\sim2\farcs9$ in right ascension and $\sim2\farcs2$ in declination, providing a crude internal consistency check.

These intersections yield our adopted cluster-center estimate as their mean:
\begin{equation*}
(\alpha,\delta)_{\rm WW} = (197.87560,-1.34195).
\end{equation*}

A somewhat similar result of $(197.87532,-1.34276)$ was found without pruning any of the quartets. The $\sim3.1\arcsec$ offset is mainly in the opposite direction of other centers, such as X-ray and SZ centers. We repeated the entire Wynne–Witt procedure using a set of image coordinates which coincide more closely with the observed centroids in the HST frames, but the catalog omits several systems that we analyse here—systems that are in any case discarded by our offset and ellipticity pruning. After retaining the corresponding clean quartets, this data yields a slightly more dispersed trio of Witt–hyperbola intersections yet a similar estimate for the center: $(\alpha,\delta)^{\rm Lim}_{\rm WW} = (197.87547,\,-1.34201)$. This agreement strengthens the robustness of our center estimate to modest changes in the underlying image measurements.

\section{Comparison with Other Centers} \label{sec:compare}

\subsection{Gas-Based Tracers}
We compare $(\alpha,\delta)_{\rm WW}$ with independent centers derived from the XMM X-ray surface-brightness peak and the AMiBA SZ decrement (Figure~\ref{fig:gas_compare}). The positional offsets are:
\begin{align*}
\Delta\mathbf x_{\text{X-ray}}
  &= (-11\farcs16,\, +1\farcs97),\quad  % thin space inside the pair, thick after comma
     \left|\Delta\mathbf x_{\text{X-ray}}\right| = 11\farcs34\\
\Delta\mathbf x_{\mathrm{SZ}}
  &= (-10\farcs42,\, +2\farcs91),\quad
     \left|\Delta\mathbf x_{\mathrm{SZ}}\right| = 10\farcs82.
\end{align*}
The Wynne--Witt center thus lies within $\sim11\arcsec$ of both gas-based tracers. 

The Wynne--Witt (WW) center is governed by the \emph{total} mass distribution on scales comparable to the Einstein ring, whereas the X--ray peak and SZ maximum trace the \emph{thermodynamic state} of the intracluster medium (ICM) in the core (X--ray emissivity $\propto n_e^2$, SZ Compton-$y$ $\propto \int n_e T_e\,dl$). These tracers can therefore identify different points in projection, as established in classic reviews of X--ray ICM physics and of the thermal SZ effect \citep{Sarazin1988, BoehringerWerner2010, Birkinshaw1999, Carlstrom2002}.

A triaxial dark-matter halo can project a stable mass centroid on the annulus near the Einstein radius while the gas peak responds to pressure and density gradients within the central few tens of kpc; radial variations in ellipticity or position angle further separate centers defined at different weighting scales. For Abell~1689, independent multi-probe analyses support a triaxial configuration with a substantial line-of-sight elongation \citep{OguriBlandford2009, SerenoUmetsu2011, Morandi2011, Umetsu2015}, providing a geometric context for center differences.

ICM dynamics also shape the gas-based centers. Minor interactions can induce ram-pressure displacements and long-lived sloshing that maintain offsets between the X--ray peak and the potential minimum \citep{AscasibarMarkevitch2006, MarkevitchVikhlinin2007}. Non-thermal pressure (turbulence, bulk motions, magnetic fields) partially supports the gas and can shift the locations of maximum pressure and of maximum density-squared relative to the potential minimum \citep{LauKravtsovNagai2009, NelsonLauNagai2014}. Central feedback from the brightest cluster galaxy creates cavities and shocks that perturb the core pressure and density structure \citep{McNamaraNulsen2007, Fabian2012}.

For Abell~1689 specifically, the WW center derived from multiple clean quartets indicates a smooth, elliptical mass configuration on strong-lensing scales, while the $\sim$11\arcsec\ west--northwest displacements of the X--ray and SZ centers (relative to WW) indicate a baryonic core which is in a slightly non-equilibrium position relative to the halo or in equilibrium with small scale structure that WW doesn’t sample . In addition to the magnitudes of the displacements, the \emph{directions} of the X--ray and SZ offset vectors relative to the lensing major axis carry diagnostic weight. Comparing their position angles with the ellipse-derived major-axis direction tests whether the baryonic displacement tracks the cluster's global elongation (consistent with a triaxial, anisotropic halo) or is driven by more local core processes (e.g., sloshing or feedback). For Abell~1689, both gas-based offsets point toward the west--northwest and differ by only $\sim$20--26$^\circ$ from the lens major axis ($P.A.\approx100$--104$^\circ$, modulo 180$^\circ$), indicating a partial alignment with the cluster's large-scale elongation.

\subsection{Lensing-Based Centers}
 
We also compare $(\alpha,\delta)_{\rm WW}$ with two lensing-based references: (i) the \emph{central} flexion peak reported by \citet{Okura2008}, which they identify with the cluster’s BCG and (ii) the strong-lensing total-mass center from \citet{Limousin2007} at $(197.87333,\,-1.34102)$. \citet{Okura2008} do not tabulate explicit coordinates for the central peak, stating that there were not sufficient background objects near the center to select a singular center. However, they do state it coincides with the BCG. We adopt the BCG position from the SIMBAD Astronomical Database, $(197.87292,\,-1.34111)$, to enable a direct numerical comparison.

\begin{deluxetable}{lcccc}[t]
\tablecaption{Offsets relative to $(\alpha,\delta)_{\rm WW}$
\label{tab:lensing_centers}}
\tablewidth{0pt}
\tablehead{
\colhead{Reference} & \colhead{Approach} & \colhead{$\Delta$R.A.} & \colhead{$\Delta$Dec.} & \colhead{$|\Delta\mathbf x|$} \\
\colhead{} & \colhead{} & \colhead{(\arcsec)} & \colhead{(\arcsec)} & \colhead{(\arcsec)}
}
\startdata
\makecell[l]{\citeauthor{Okura2008}\\(\citeyear{Okura2008})}   
  & Flexion 
  & $-9.66$ & $+3.02$ & $10.12$ \\
\makecell[l]{\citeauthor{Limousin2007}\\(\citeyear{Limousin2007})} 
  & Strong Lensing 
  & $-8.17$ & $+3.36$ & $8.84$ \\
\enddata
\end{deluxetable}

The central flexion peak (\citet{Okura2008}, anchored to the SIMBAD BCG coordinates) lies $(-9\farcs66,\,+3\farcs02)$ from $(\alpha,\delta)_{\rm WW}$, with $|\Delta\mathbf x|=10\farcs12$. The strong-lensing mass-model center from \citet{Limousin2007} lies $(-8\farcs17,\,+3\farcs36)$ away, with $|\Delta\mathbf x|=8\farcs84$. The close agreement of these independent cluster-scale lensing tracers with the Wynne--Witt center supports a coherent mass centroid on strong-lensing scales.

\section{Discussion of Results} \label{sec:discuss}

The Wynne--Witt application to Abell~1689 yields a stable lensing center, retained quartets with consistent geometry, and $\sim$11\arcsec\ offset with gas-based tracers, indicating a largely smooth elliptical mass distribution with mild baryonic displacements.

\begin{enumerate}[noitemsep]
  \item Three quartets (4, 9, 8) meet the $|\Delta\mathbf x_{WW}|\le6\farcs25$ consistency threshold and exhibit mutually consistent ellipse parameters ($P.A.\approx100$--$104^\circ$, $a/b\approx1.2$--$1.5$), whereas the rejected quartets show larger offsets and/or inconsistent parameters (Table~\ref{tab:quads}).

  \item The pairwise intersections of the three Witt hyperbolae give a lensing center of $(197.87560,-1.34195)$ (J2000), with the intersections clustering within $\sim2\farcs9$ in R.A. and $\sim2\farcs2$ in Dec.

  \item Relative to gas-based tracers, the lensing center lies $11\farcs34$ from the X-ray peak and $10\farcs82$ from the SZ maximum with offset vectors $\Delta\mathbf x_{\rm X\text{-}ray}=(-11\farcs16,+1\farcs97)$ and $\Delta\mathbf x_{\rm SZ}=(-10\farcs42,+2\farcs91)$ (Fig.~\ref{fig:gas_compare}). The corresponding position angles differ by $\sim20$--$26^\circ$ from the lens major axis ($P.A.\approx100$--$104^\circ$), indicating partial alignment with the cluster’s large-scale elongation.

  \item Relative to other lensing tracers, the central flexion peak of \citet{Okura2008} is $(-9\farcs66,\,+3\farcs02)$ from $(\alpha,\delta)_{\rm WW}$ ($|\Delta\mathbf x|=10\farcs12$), while the strong-lensing mass-model center of \citet{Limousin2007} is $(-8\farcs17,\,+3\farcs36)$ away ($|\Delta\mathbf x|=8\farcs84$). We do not compare to weak-lensing centers for the reasons stated in Sections~\ref{sec:intro} and~\ref{sec:compare}.

  \item An independent image catalog gives a slightly larger intersection dispersion and a mean of $(197.87547,\,-1.34201)$, consistent with the adopted center.
\end{enumerate}

\section{Opportunities for further work} \label{sec:future_work}
There are several extensions that would sharpen uncertainties and broaden applicability of the Witt-Wynne construction:
\begin{enumerate}[noitemsep]
  \item Propagate image-position errors through the conic construction; bootstrap resample quartets; and assess sensitivity to the pruning threshold. Quote a covariance and error ellipse for $(\alpha,\delta)_{\rm WW}$.
  \item Systematically compare WW centers with weak-lensing convergence peaks and flexion peaks as a function of aperture/scale, including vector-angle diagnostics (alignment versus transverse displacements).
  \item Quantify how X--ray/SZ peaks move with smoothing kernel and aperture, and relate the offset vectors to morphological indicators (e.g., concentration, sloshing signatures, cavity locations).
  \item Automate the cross-check against alternative image catalogs and report center shifts; weight hyperbola intersections by $1/|\Delta\mathbf x_{WW}|$ and evaluate alternatives.
  \item Apply the workflow to additional lens-rich clusters to map the distribution of lensing-vs-gas offset amplitudes and directions across dynamical states.
\end{enumerate}

\appendix

\section{Construction of Witt's Hyperbola}\label{app:hyperbola}

In contrast to the approach taken by \citet{Wynne}, we work in a 2D projection of sky coordinates, rather than a rotated system in which coordinate axes are the asymptotes of the hyperbola. Beginning with the general form for a hyperbola and applying the rectangular hyperbola constraint of $C_H=-A_H$ and an overall scale chosen by $F_H\equiv+1$, we find the form of a rectangular hyperbola with a given scale:
\begin{equation}
Q_H(x,y)=A_H(x^2-y^2)+B_Hxy+D_Hx+E_Hy+1=0,
\qquad B_H^{2}+4A_H^2=0.\label{eq:hyperbola_conic}
\end{equation}

\subsection{Coefficient determination from four images.} 

Let \(x_i,y_i\) denote the four image coordinates of a single quadruply--lensed source. From the rectangular constraint and the overall scale, we are left with four unknowns $(A_H,\,B_H,\,D_H,\,E_H)$. Enforcing that each image satisfies $Q_H(x_i,y_i)=0$ yields the linear system
\begin{equation}
\begin{bmatrix}
 x_1^{2}-y_1^{2} & x_1y_1 & x_1 & y_1\\
 x_2^{2}-y_2^{2} & x_2y_2 & x_2 & y_2\\
 x_3^{2}-y_3^{2} & x_3y_3 & x_3 & y_3\\
 x_4^{2}-y_4^{2} & x_4y_4 & x_4 & y_4
\end{bmatrix}
\begin{pmatrix} A_H\\ B_H\\ D_H\\ E_H \end{pmatrix}
 =-\begin{pmatrix}1\\1\\1\\1\end{pmatrix},
\label{eq:hyperbola_linear_system}
\end{equation}
where the right--hand side arises from moving the fixed constant term $F_H=+1$ to the right. The $4\times4$ matrix in Eq.~\ref{eq:hyperbola_linear_system} is a Vandermonde--like matrix whose nonsingularity requires that the four points be non--collinear and mutually distinct; in that case the system admits a unique solution and the full coefficient set $(A_H,\,B_H,\,C_H=-A_H,\,D_H,\,E_H,\,F_H=1)$ is obtained.

\section{Construction of Wynne's Ellipse}\label{app:ellipse}

Throughout we assume $C\neq0$, so the asymptotes are not aligned with the coordinate axes. Additionally, we characterize a conic's orientation by the slopes of its axes rather than by a position angle to avoid the use of trigonometric functions.

\begin{enumerate}[label=(\roman*)]
\item \textbf{Asymptote slopes.}  
      The quadratic part $A_Hx^{2}+B_Hxy+C_Hy^{2}$ factors into two linear forms whose slopes are
      \begin{equation}
        m_{1,2}= \frac{-B_H\pm\sqrt{\,B_H^{2}-4A_HC_H\,}}{2C_H}.
      \end{equation}

\item \textbf{Ellipse subject to a fixed orientation.}  
      For each slope $m\in\{m_1,m_2\}$ we determine the conic
      \begin{equation}
        Q_E(x,y)=A_E x^{2}+B_E xy+C_E y^{2}+D_E x+E_E y+F_E=0
      \end{equation}
      that (a) passes through the four images and (b) obeys the orientation constraint
      \begin{equation}
        B_E - k\bigl(A_E-C_E\bigr)=0,
        \qquad k=\frac{2m}{1-m^{2}}.
      \end{equation}
      These five homogeneous constraints form a $5\times6$ linear system whose one-dimensional null-space fixes the ellipse coefficients up to scale.

\item \textbf{Ellipse center and branch selection.}  
      The center $(x_e,y_e)$ of each candidate ellipse is obtained from
      \begin{equation}
        \begin{bmatrix}
          2A_E & B_E \\ B_E & 2C_E
        \end{bmatrix}
        \begin{pmatrix}x_e\\y_e\end{pmatrix}
        =-\begin{pmatrix}D_E\\E_E\end{pmatrix}.
      \end{equation}
      Its distance to the hyperbola, measured along the normal, is
      \begin{equation}
          d=\min_{(x,y)\in Q_H} \sqrt{(x-x_e)^{2}+(y-y_e)^{2}},
      \end{equation}
      which can be expressed via a Lagrange-multiplier parameter $\lambda$. Eliminating $x$ and $y$ reduces the stationarity conditions to a single polynomial equation $P(\lambda)=0$ of degree at most six. Solving $P$ for its real roots supplies all candidate distances; the minimal one is adopted. The ellipse whose center yields this smallest $d$ is taken as Wynne’s ellipse, with center $\mathbf c_W=(x_e,y_e)$.
\end{enumerate}

\section{Comparison with Weighted Best-Fit Schemes}\label{app:comparison}

\citet{Luhtaru} developed a scheme for fitting Witt-Wynne models to quadruple images that used a \emph{weighted least‑squares} approach: rather than enforce the four images as exact roots of both conics, they choose hyperbola–ellipse pairs that minimize a cost function built from the images’ orthogonal distances to the curves. The present work instead imposes \emph{exact incidence}—the conics pass through every image—and evaluates self‑consistency \emph{post‑facto} via $|\Delta\mathbf x_{WW}|$, defined by the offset between Witt's Hyperbola and the center of Wynne's Ellipse. Here we summarize the principal differences and their practical consequences.

\begin{description}[leftmargin=1.5em,itemsep=0em]

 \item[Construction paradigm]Our method is \emph{deterministic}. Solving Eq.~\ref{eq:hyperbola_linear_system} and the corresponding ellipse constraints yields closed‑form coefficients in constant time without iteration. In contrast, the weighted scheme solves a non‑linear minimization problem by iteration whose convergence depends on the chosen weight prescription and the initial guesses for the conic coefficients.

 \item[Handling of tight image pairs]The least‑squares approach must guard against configurations in which Witt’s hyperbola and Wynne’s ellipse are nearly tangent. In such cases two images lie close together on the secondary branch. To that end, it multiplies each image’s squared distance by $w_i=\min\bigl[10^{0.5(d_{\max}/d_{\min})^{2}},1000\bigr]$. Our analytic construction automatically places both close images \emph{exactly} on the conics, and any residual inconsistency appears in $|\Delta\mathbf x_{WW}|$.

 \item[Computational cost]Per quartet, the analytic route requires a single small matrix inversion and a degree‑6 polynomial root—negligible even for hundreds of quartets. The weighted fit incurs higher cost if one seeks full non‑linear optimization, though for their sample the overhead remained acceptable.

 \item[Bias versus variance]Imposing exact position matching eliminates one source of model variance but risks \emph{bias} if real cluster potentials deviate substantially from ellipticity; such quartets are then discarded, potentially reducing statistical power. The least‑squares method retains more data but can suffer from bias if the weighting scheme disproportionately down‑weights informative yet widely separated images.
\end{description}

In summary, the analytic Wynne–Witt construction adopted here offers speed, conceptual transparency, and a crisp geometric diagnostic in $|\Delta\mathbf x_{WW}|$. Weighted best‑fit schemes provide greater elasticity at the price of additional hyper‑parameters (the weight rule) and the need for iterative optimization. In practice, both yield comparable center estimates for clean quartets; differences emerge chiefly in how each method treats non-elliptical potential configurations.

\bibliography{main}{}
\bibliographystyle{aasjournalv7}

\end{document}